\documentclass[12pt]{article}
\usepackage[T2A]{fontenc}
\usepackage{graphicx}
\usepackage[cp1251]{inputenc}
\begin{document}

\newcommand{\Teff}{$T_{\rm eff}$}
\newcommand{\lgg}{log\,$g$}
\newcommand{\eps}[1]{\log\varepsilon_{\rm #1}}
\newcommand{\kms}{km\,s$^{-1}$}
\newcommand{\kH}{$S_{\!\rm H}$}    %%% Note negative spaces!
\newcommand{\Eexc}{$E_{\rm exc}$}
\newcommand{\iso}[1]{\mbox{$^{#1}{\rm Ba}$}}
\newcommand{\eu}[5]{\mbox{$#1\,^#2{\rm #3}^{#4}_{\rm #5}$}}

\def\physscr{Physica Scripta}
\def\aaps{A\&A Suppl. Ser.}
\def\mnras{MNRAS}
\def\apj{ApJ}
\def\aap{A\&A}

\centerline{\bf Even-to-Odd Barium Isotope Ratio in Selected Galactic Halo Stars}

\bigskip
\centerline{Lyudmila Mashonkina$^*$\footnote{\tt E-mail: lima@inasan.ru}, Andrey K. Belyaev$^{**}$\footnote{\tt E-mail: andrey.k.belyaev@gmail.com}}

\bigskip
\centerline{\it $^*$Institute of Astronomy, Russian Academy of Sciences,}
\centerline{\it Pyatnitskaya st. 48, 119017 Moscow, Russia}
 
\centerline{\it $^{**}$Department of Theoretical Physics and Astronomy, Herzen University,}
\centerline{\it Moika st. 48, St. Petersburg 191186, Russia }
 
\bigskip
{\bf Abstract -}
We have updated the Ba~II model atom by taking into account the H~I impact excitation with the rate coefficients from the quantum-mechanical calculations
of Belyaev and Yakovleva (2018). Using high-resolution stellar spectra and the non-local thermodynamic equilibrium (non-LTE) line formation for Ba~II, we have determined the fraction of barium isotopes with an odd mass
number ($f_{\rm odd}$) in four Galactic halo giants with well-known atmospheric parameters. We use a method
based on the requirement that the abundances from the Ba~II 4554\,\AA\ resonance and Ba~II 5853, 6496\,\AA\ subordinate lines be equal. An accuracy of 0.04~dex in determining the barium abundance from individual
lines has been achieved. In three stars (HD~2796, HD~108317, and HD~122563) $f_{\rm odd} \ge $ 0.4. This suggests that $\ge$ 80~\%\ of the barium observed in these stars was synthesized in the $r$-process. In HD~128279 $f_{\rm odd}$ = 0.27 exceeds the fraction of odd barium isotopes in the Solar system, but only slightly. The dominance of the $r$-process at the formation epoch of our sample stars is confirmed by the presence of a europium
overabundance relative barium, with [Eu/Ba] $>$ 0.3. We have calculated the non-LTE abundance corrections for five Ba~II lines and investigated their dependence on atmospheric parameters
in the ranges of effective temperatures from 4500 to 6500~K, surface gravities log~$g$ from 0.5 to 4.5, and
metallicities [Fe/H] from 0 to $-3$.

Keywords: {\it stellar atmospheres, spectral line formation, excitation and charge exchange in inelastic collisions with H~I atoms, barium odd-to-even isotope abundance ratios in stars}

\section{Introduction}

The isotopes of the elements located in Mendeleyev's periodic table beyond the iron group ($Z >$ 30, hereafter, heavy elements) are synthesized in neutron-capture nuclear reactions (Burbidge et al. 1957), which are subdivided into slow ($s$-) and rapid ($r$-) process of neutron-captures, depending on the neutron flux density. The
$s$-process is divided into two components, depending on its site and the production efficiency of isotopes of
various masses ($A$). The main component is associated with the thermally pulsing asymptotic giant branch (AGB) phase of intermediate mass stars and the production of isotopes with $A$ = 90-208. As confirmed by theoretical and observational studies (for a review, see Busso et al. 1999), stars with initial masses of
1-4 $M_\odot$ made the greatest contribution to the solar abundance of $s$-nuclei. The weak component runs
in massive stars ($> 10M_\odot$) at the core helium burning stage and produces the lightest isotopes of heavy elements, with $A \le$ 90 (see, e.g., K\"appeler et al. 1989).

The $r$-process overcomes the barrier near bismuth ($Z$ = 83, $A$ = 208) and can synthesize the heaviest nuclei. It is probably associated with more than one type of stars. The models of neutron star or black hole mergers, type II supernova explosions, accretion-driven gravitational collapse, etc. are proposed in the literature, and so far there is no general consensus (Nishimura et al. 2017). However, all agree that the $r$-process must occur in stars more massive than those in which nucleosynthesis proceeds in the main component of the $s$-process. The point is that different elements and different isotopes of the same element are synthesized with different efficiencies in the $s$- and $r$-processes. For example, in Solar system matter 80\,\%\ of the barium are the $s$-nuclei synthesized in AGB stars, while europium is a product of predominantly the $r$-process, by 94\,\%\ (Travaglio et al. 1999). Barium ($A$ = 134-138) and europium ($A$ = 151, 153) cannot be produced in the weak component of the $s$-process. Back in the late 1970s Galactic halo stars were shown to exhibit europium
overabundances relative to barium relative to the corresponding solar ratio, i.e., [Eu/Ba]\footnote{We use the standard notion for elemental ratios: [X/Y] = log$(N_X/N_Y)_{*}$  -- log$(N_X/N_Y)_{\odot}$.} $> 0$ (Spite
and Spite 1978). This is possible only if the $r$-process proceeds in stars with a shorter evolution time, i.e., more massive than AGB stars. Theoretical arguments for the dominance of the $r$-process over the $s$-process at the formation epoch of the halo stellar population were advanced by Truran (1981).

To solve the problem of the astrophysical site for the $r$-process and to refine the present views of the chemical evolution of the Galaxy, it is very important to reconstruct the history of enrichment of Galactic
matter with $s$- and $r$-nuclei. The [Eu/Ba] ratio is a good indicator of the ratio of the contributions
from the $r$- and $s$-processes ($r/s$) to the barium abundance at the epoch when the star was formed. If the $r$-process dominated, then the star has [Eu/Ba] = [Eu/Ba]$_r$, which in different works on calculating
the contribution from the main component of the $s$-process to the solar europium and barium abundances varies from 0.67 (Travaglio et al. 1999) to 0.80 (Bisterzo et al. 2014) and is 0.63 in the theoretical waiting-point (WP) approximation (Kratz et al. 2007). Mashonkina et al. (2003) showed than in thick-disk and halo stars [Eu/Ba] lies in the range from 0.35 to 0.67 and nucleosynthesis in AGB stars had already begun at the formation epoch of the thick disk. The contribution of the $s$-process to the barium abundance was estimated to be from 30 to 50\,\%, and the thick disk was shown to have been formed in the
interval between 1.1 and 1.6~Gyr from the onset of protogalactic collapse. Having analyzed the ratios
between Ba, La, Nd, Eu, and Dy, Burris et al. (2000) concluded that the contribution of AGB stars becomes
noticeable starting from [Fe/H] = $-2.3$. Simmerer et al. (2004) lowered this boundary to [Fe/H] = $-2.6$ by analyzing the La/Eu ratio, while Roederer et al. (2010), on the contrary, raised it to [Fe/H] = $-1.4$ based on Pb/Eu observations.

Another indicator of $r/s$ in the barium abundance is the fraction of isotopes with an odd mass number, $f_{\rm odd}$. In Solar system matter barium is represented mainly by five isotopes with the following fraction in
the total abundance: \iso{134} : \iso{135} : \iso{136} : \iso{137} : \iso{138} = 2.4 : 6.6 : 7.9 : 11.2 : 71.7 and $f_{\rm odd}$ = 0.18 (Lodders et al. 2009). A different mixture of isotopes is produced in the $r$-process; for example, Travaglio et al. (1999) predicted \iso{135} : \iso{137} : \iso{138} = 24 : 22 : 54 and $f_{\rm odd,r}$ = 0.46. A close value is obtained in the WP model: $f_{\rm odd,r}$ = 0.438 (Kratz et al. 2007). The
Ba~II 4554 and 4934\,\AA\ resonance lines are sensitive to a variation of the mixture of barium isotopes, and this makes it possible to determine $f_{\rm odd}$ and, hence, $r/s$ in stars.

Two methods of determining $f_{\rm odd}$ are used in the literature. Both are based on the fact that for the
isotopes \iso{135} and \iso{137} there is hyperfine splitting (HFS) of levels and each barium line consists of a set of components. The first method uses the fact that the width of the Ba~II 4554\,\AA\ resonance line increases with $f_{\rm odd}$. This method requires a very high spectral resolution ($R$) and a superhigh signal-to-noise ratio ($S/N$). For example, for the nearest ($d$ = 62 pc) halo star, HD~140283, Lambert and Allende Prieto (2002) took a spectrum with $R\simeq 200\,000$ and $S/N\simeq 550$, while Gallagher et al. (2010) took a spectrum with $S/N\simeq 1\,110$ (!). The complexity of the method is related to the necessity of separating the line broadening due to the hyperfine structure from the broadening by rotation, macroturbulent motions in the stellar atmosphere, and the instrumental broadening. In the classical approach (homogeneous and 1D model atmospheres) lines in adjacent spectral regions, as a rule, of iron-group elements are used for this purpose. The weakness of this approach is obvious. Whereas the instrumental and rotational broadenings produce
the same effect on lines with close wavelengths, the effect of macroturbulence can depend on the line
formation depth. It cannot be properly taken into account for the Ba~II resonance lines using the (predominantly subordinate) lines of other elements. That is why the results of different authors for the same star
differ so much. For example, for HD~140283 Lambert and Allende Prieto (2002) obtained $f_{\rm odd}$ = 0.3$\pm$0.21, which points to a significant contribution of the $r$-process to the barium abundance, while Gallagher
et al. (2010) gave $f_{\rm odd}$ = 0.02, which is lower than that in a pure $s$-process: $f_{\rm odd,s}$ = 0.11 (Arlandini et al. 1999). The results obtained by Gallagher et al. (2015) with the application of hydrodynamic
calculations in a 3D model atmosphere are encouraging: $f_{\rm odd}$ = 0.38 for the same star, which is consistent with the present views of the history of heavy-element production in the Galaxy. However, it is unlikely that the method will be widely used, because it is impossible to obtain the observed spectra of the required
quality for metal-poor stars, which are mostly at great distances.

The second method uses the fact that the HFS effect is different for the Ba~II resonance and subordinate
lines. First, the total barium abundance is determined from the subordinate lines and then $f_{\rm odd}$
is varied until the same abundance is deduced from the resonance lines. The idea was proposed by Magain
and Zhao (1993), but it was used very rarely. The method does not require a superhigh quality of
the observed spectrum, but it requires an accurate determination of the atmospheric parameters, especially
the microturbulence ($\xi_t$), and abandoning the LTE assumption in Ba~II line calculations, because
the non-LTE effects are different for the resonance and subordinate lines. Therefore, the method was
applied for a small number of stars and predominantly in our papers: Mashonkina and Zhao (2006, 25 stars), Mashonkina et al. (2008, 2 stars), and Jablonka et al. (2015, 2 stars).

The goal of this paper is to determine the relative contribution of the $r$- and $s$-processes to the barium
abundance in four halo giants by determining the fraction of odd barium isotopes from an analysis of
the Ba~II resonance and subordinate lines. Our study is motivated by two factors. First, the parallaxes
measured by the Gaia observatory were published in April 2018 (Gaia Collaboration 2018), which allow
accurate surface gravities (log~$g$) to be obtained for stars up to $\sim$2.5 kpc away. Second, quantum-mechanical calculations of Ba~II + H~I collisions were performed for the first time in 2018 (Belyaev and
Yakovleva 2018). Using these data in non-LTE calculations will allow the degree of confidence in the
results obtained to be increased.

\begin{table*}  %[htbp]
\caption{Atmospheric parameters of the selected stars and characteristics of observed spectra. $\xi_t$ is in \kms.}
\label{tab:param}
\begin{center}
	\begin{tabular}{rcccclcl}
		\hline\hline
		\multicolumn{1}{l}{HD}  & \Teff & \lgg & [Fe/H] & $\xi_t$ & \multicolumn{3}{c}{Spectra} \\
		\cline{6-8}
		  &  K    &      &        &    & Spectrograph & R & Ref. \\
		\hline      
		  2796 & 4880 & 1.80$\pm$0.03 & $-2.19$ & 1.8 & VLT2/UVES & 70\,000 &  ID: 076.D-0546(A) \\
		108317 & 5270 & 2.81$\pm$0.02 & $-2.24$ & 1.4 & Magellan/MIKE & 60\,000 & Ezzeddine (2017) \\
		122563 & 4600 & 1.40$\pm$0.02 & $-2.57$ & 1.6 & VLT2/UVES & 80\,000 & Bagnulo et al. (2003) \\
		128279 & 5200 & 3.00$\pm$0.01 & $-2.19$ & 1.1 & VLT2/UVES & 45\,000 &  ID: 71.B-0529(A) \\
\noalign{\smallskip}\hline \noalign{\smallskip}
	\end{tabular}
\end{center}
\end{table*}

The sample of stars, observational data, and atmospheric parameters are presented in Section~1. The
non-LTE calculations for Ba~II are presented in Section~2. We determine the barium and europium abundances
in Section~3 and the fraction of odd barium isotopes in Section~4. Our conclusions are formulated
in Section~5.

\section{SAMPLE OF STARS, OBSERVATIONAL DATA, AND ATMOSPHERIC PARAMETERS}\label{Sect:stars}

The formulated problem requires a very high accuracy of determining the barium abundance from
individual lines. Therefore, we exclude the blended Ba~II 4934 and 6141\,\AA\ lines and use three lines:
one resonance, Ba~II 4554\,\AA, and two subordinate, Ba~II 5853 and 6497\,\AA, lines. To choose the stars,
we analyzed the lists from Zhao et al. (2016) and Mashonkina et al. (2017). We were guided by the
following criteria:

\begin{enumerate}
	\item The Ba~II 4554\,\AA\ line should not be very strong to avoid the possible influence of the
chromospheric temperature rise in the stellar atmosphere, but it should remain sensitive to a variation in $f_{\rm odd}$. The latter implies that when changing $f_{\rm odd}$ from 0.18 (a solar mixture of isotopes) to 0.46 (a pure $r$-process), the abundance from this line should change at least by 0.1~dex.
\item Both subordinate lines should be reliably measured in the spectrum.
\item The accuracy of log~$g$ is no less than 0.05~dex.
\end{enumerate}

For various reasons, none of the stars in the list of Zhao et al. (2016) was chosen. For example, for
HD~140283 an increase in $f_{\rm odd}$ from 0.18 to 0.46 leads to a decrease in the abundance from Ba~II 4554\,\AA\ by only 0.02~dex, which is less than the error in the total abundance from the subordinate lines.

From the list of Mashonkina et al. (2017) we chose four stars satisfying all criteria (Table~1). The
Ba~II 4554\,\AA\  line in these stars has an equivalent width ($W_{obs}$) in the range from 80 to 140~m\AA. The observed spectra and effective temperatures (\Teff) were taken from the same paper. The values of log~$g$ were calculated using the distances (Bailer-Jones et al. 2018) based on the Gaia parallaxes. The formula
to calculate log~$g$ also includes \Teff, the stellar mass, the apparent $V$ magnitude, and the bolometric
correction. Since the stars are old ($-2.6 <$ [Fe/H] $\le -2.2$) and giants, their mass is reliably fixed: $M = 0.8 M_\odot$. The $V$ magnitudes were taken from the SIMBAD\footnote{http://simbad.u-strasbg.fr/simbad/} astronomical database. The bolometric corrections were calculated based on the tables from Alonso et al. (1999). The errors of log~$g$ in Table~1 correspond to the errors in the distances.

We determined the iron abundance from Fe~II lines and the microturbulence velocity from the requirement of the same abundance from Fe~I lines with different $W_{obs}$. Our non-LTE calculations for Fe~I-Fe~II were performed by the method developed previously (Mashonkina et al. 2011), which was updated by taking into account the inelastic processes in Fe~I + H~I and Fe~II + H~I collisions with the rate coefficients from
the quantum-mechanical calculations of Yakovleva et al. (2018, 2019).

\section{NON-LTE CALCULATIONS FOR Ba II}\label{Sect:method}

\subsection{The Ba II Model Atom}

In the atmospheres of metal-poor stars the electron number density is low and the inelastic processes in collisions with neutral hydrogen atoms play an important role in establishing the statistical equilibrium
(SE) of atoms. Belyaev and Yakovleva (2018) were the first to derive the rate coefficients from
their quantum-mechanical calculations of Ba~II + H~I collisions. In this paper the non-LTE calculations
for Ba~II were performed with the model atom from Mashonkina et al. (1999), which was updated by
replacing the approximate Drawinian rates of collisions with H I atoms (Drawin 1968; Steenbock and Holweger
1984) with the accurate data from Belyaev and Yakovleva (2018).

\begin{figure*}  %[htbp]
\includegraphics[width=80mm]{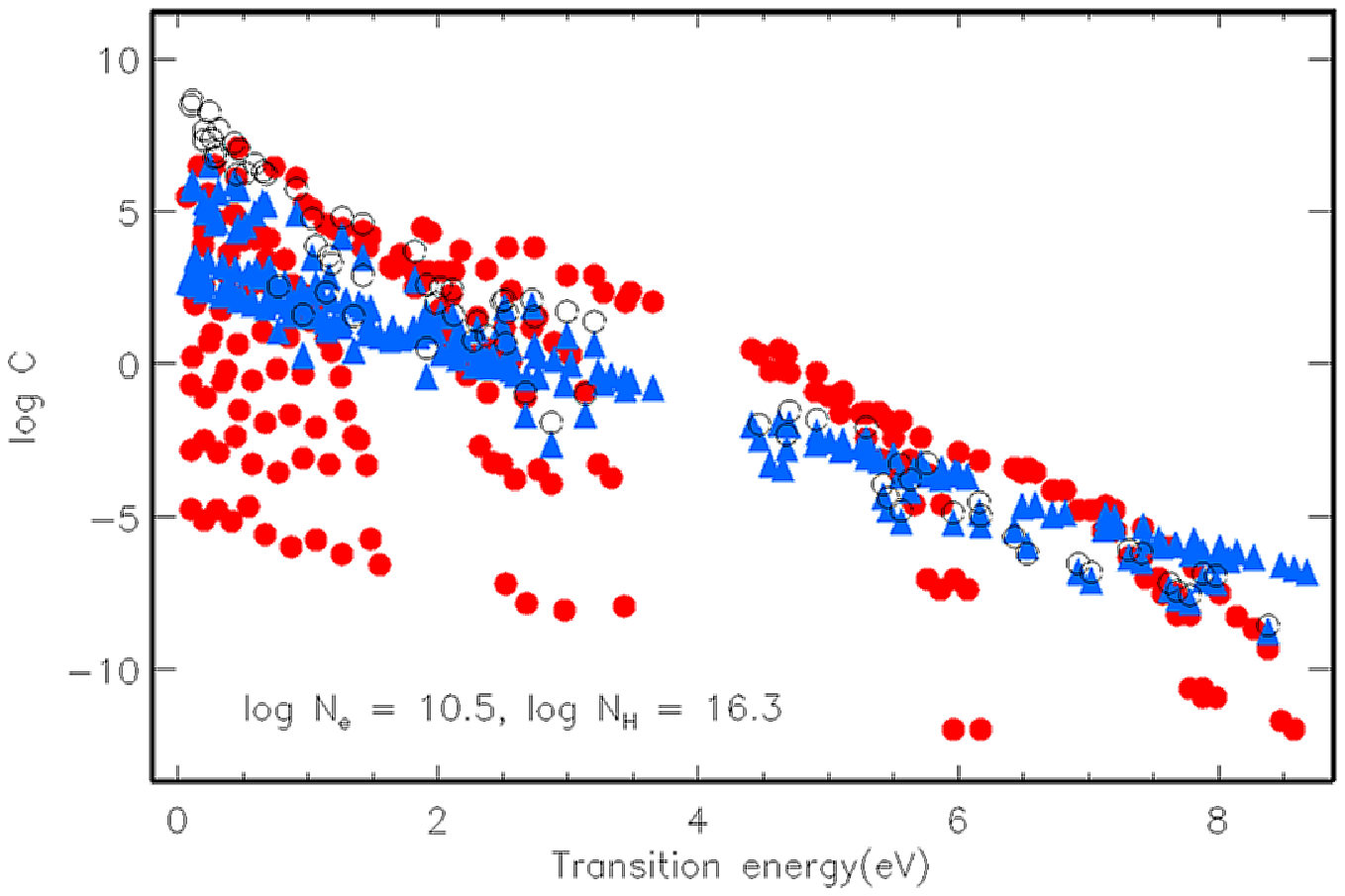}
\includegraphics[width=80mm]{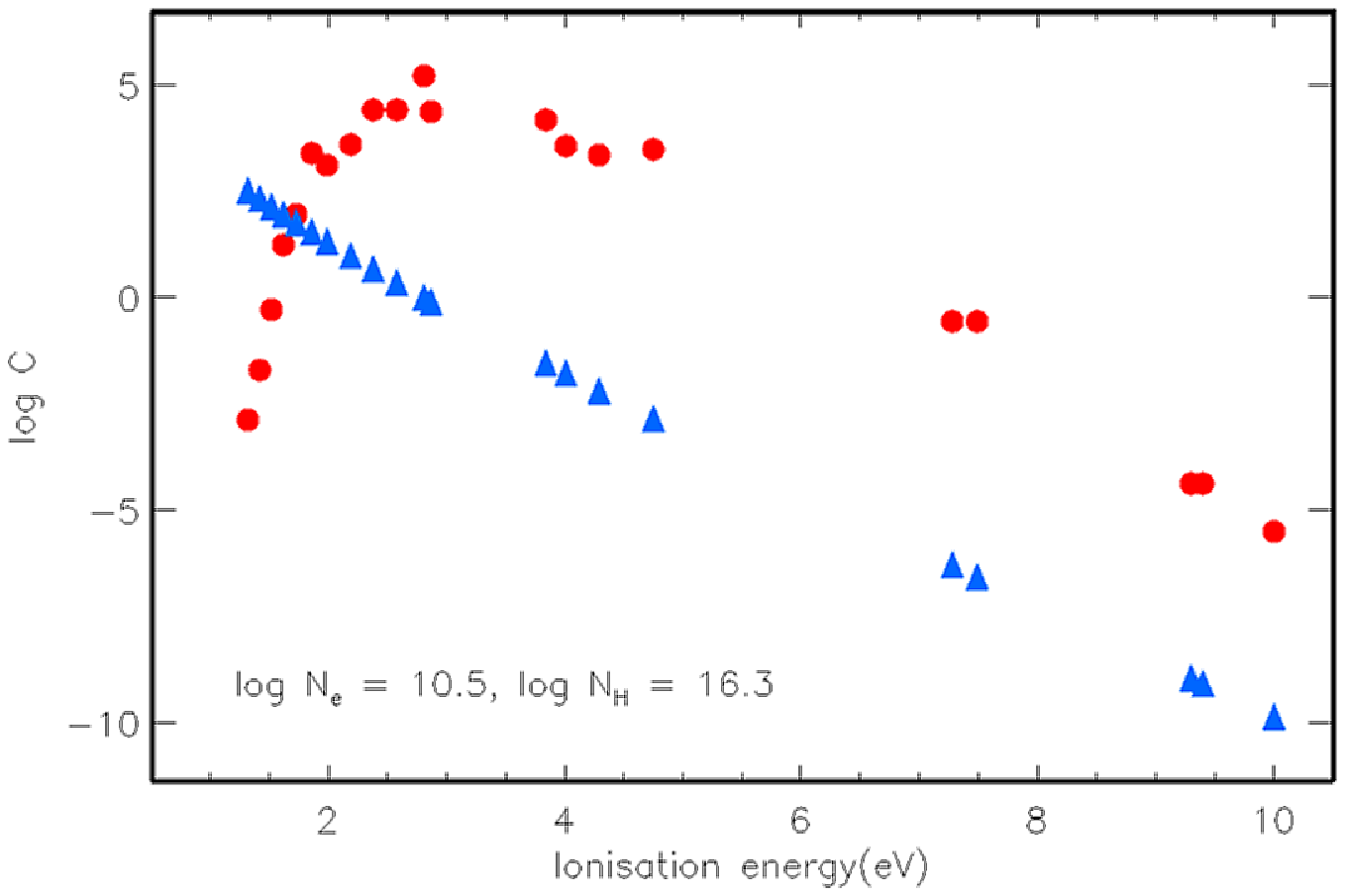}
%\hspace{-8mm} 
%\vspace{-14cm}
\caption{Left panel: Excitation rates, log~C~(СЃ$^{-1}$), for Ba~II transitions in collisions with electrons (triangles) and H~I atoms (red circles, using the data from Belyaev and Yakovleva, 2018). The open circles correspond to the Drawinian rates scaled by a factor of \kH\ = 0.01.
 Right panel: Rates of the processes Ba$^+ + {\rm e^-} \rightarrow {\rm Ba}^{++} + 2 e^-$ and Ba$^{+} + {\rm H} \rightarrow {\rm Ba^{++} + H^-}$ using similar symbols. The calculations were made for $T = 4190$~K, log~$N_{\rm e}$(cm$^{-3}$) = 10.5 and log~$N_{\rm H}$(cm$^{-3}$) = 16.3.}
\label{fig:rates}
\end{figure*}

In Fig 1a we compare the electron and H I impact excitation rates for various transitions. The calculations were made with the temperature $T$ = 4190~K, the electron number density log~$N_e$(cm$^{-3}$) = 10.5, and the H~I number density log $N_{\rm H}$(cm$^{-3}$) = 16.3, which correspond to the Ba~II line formation depth (log~$\tau_{5000} \simeq -0.5$) in the \Teff / log~$g$ / [Fe/H] = 4600~K/1.6/$-2.5$ model atmosphere. For comparison, we also provide the Drawinian rates scaled by a factor of \kH\, = 0.01 that was found empirically by Mashonkina et al. (1999). This figure shows that the collisions with H~I atoms play no lesser role in establishing
the SE of Ba~II than do the collisions with electrons, despite the fact that for transitions with close excitation energies ($E_{lu}$) the rates in the calculations of Belyaev and Yakovleva (2018) can differ by several
orders of magnitude: up to 10~dex in the energy range $E_{lu} <$ 1.5~eV important for the SE. For each level in the atom the rate of charge exchange Ba+$^{+} + {\rm H} \rightarrow {\rm Ba^{++} + H^-}$ exceeds the electron impact ionization rates by 4-5 orders of magnitude (Fig.~1b), but this does not exert a great influence on the ionization equilibrium of Ba~III/Ba~II, because Ba~II dominates in the total barium concentration.

\begin{figure*}  %[htbp]
\includegraphics[width=80mm]{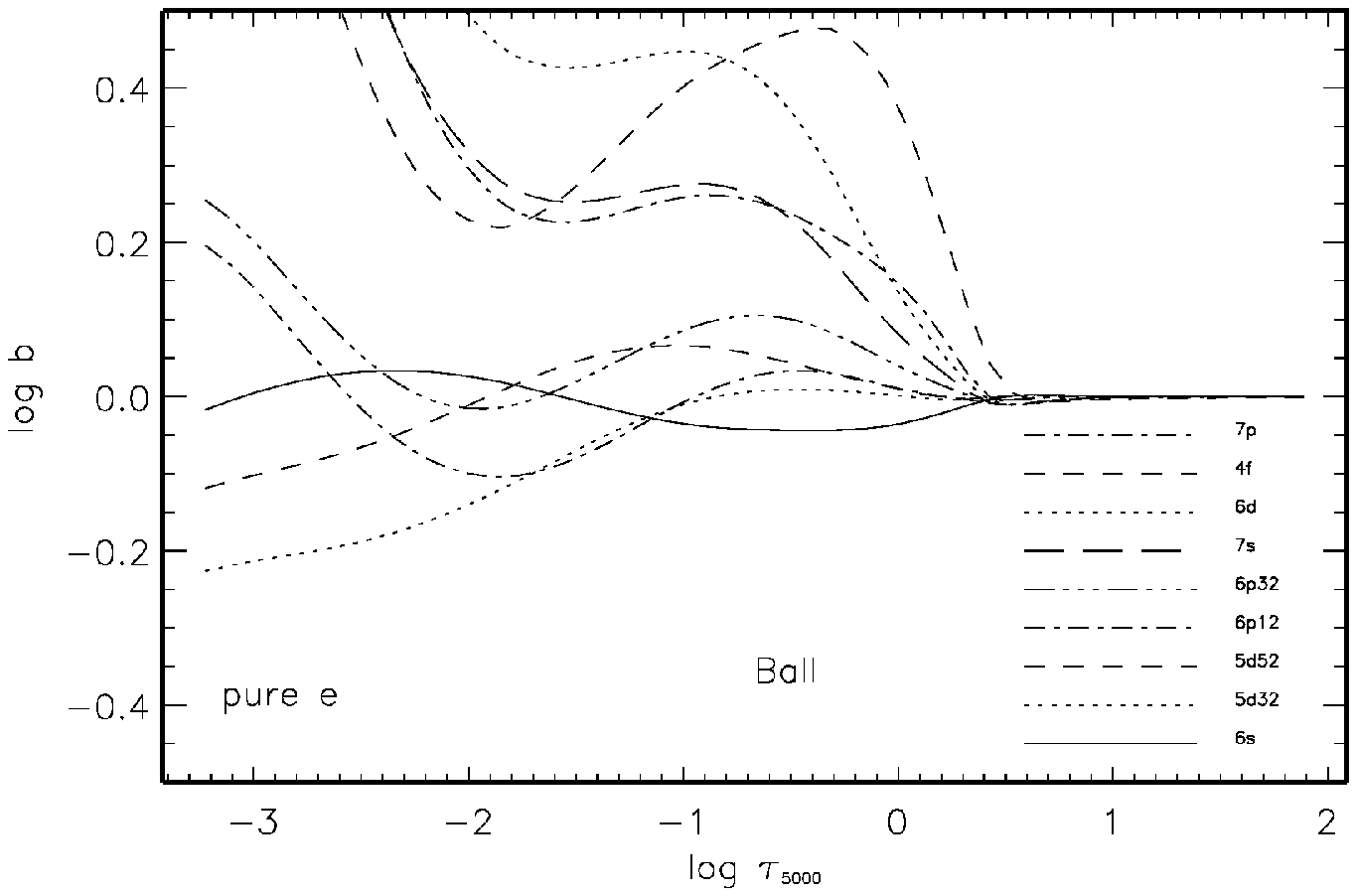}
\includegraphics[width=80mm]{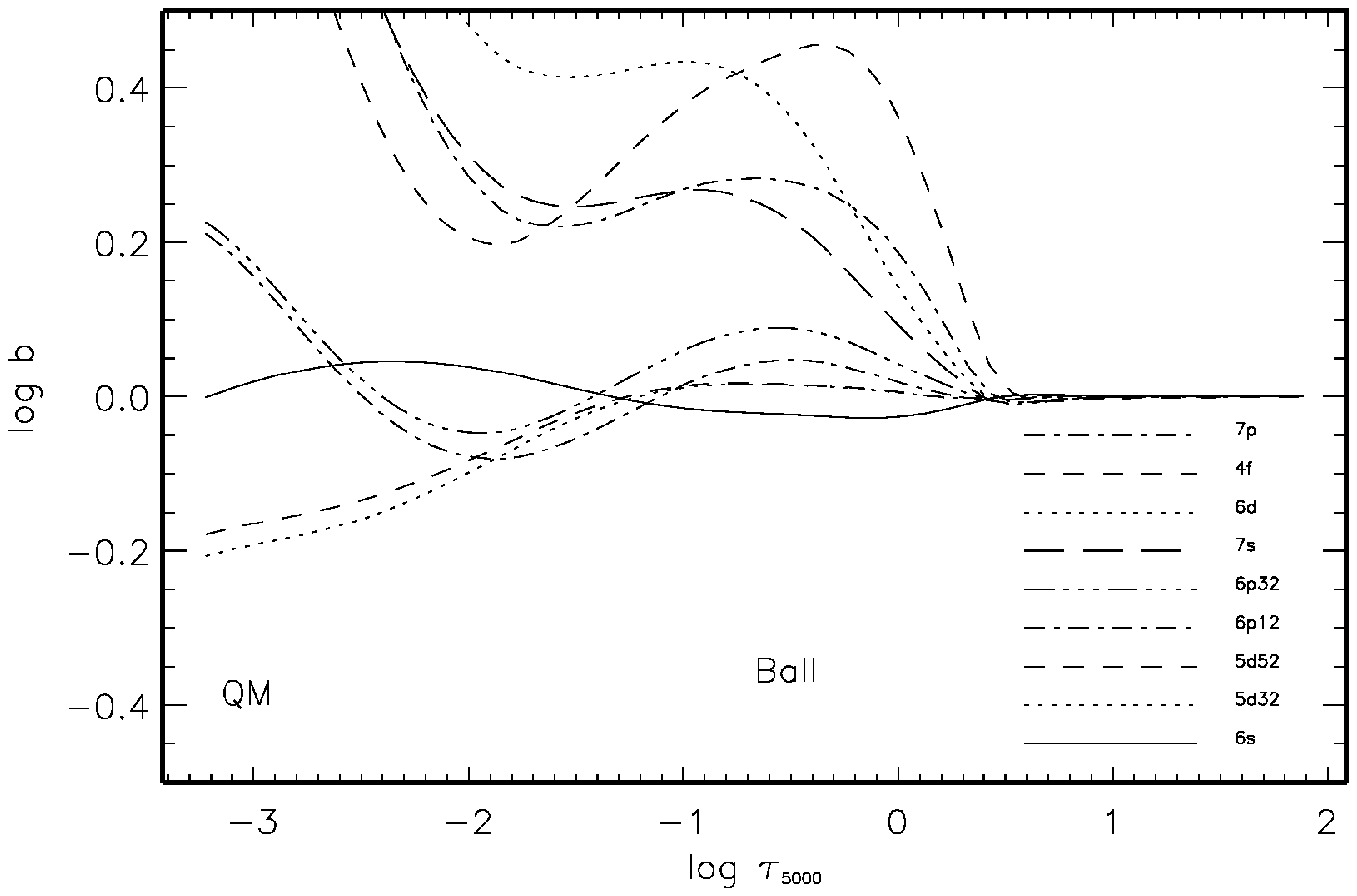}
%\hspace{-8mm} 
%\vspace{-14cm}
\caption{b-factors of selected Ba II levels in the 4600/1.40/$-2.57$ model atmosphere. Left panel: calculations with purely electron collisions. Right panel: calculations include the collisions with H~I atoms using the data from Belyaev and Yakovleva (2018). }
	\label{fig:bf}
\end{figure*}

The system of SE and radiative transfer equations was solved in a given model atmosphere using an updated DETAIL code (Butler and Giddings 1985). Figure~2 shows the b-factors, b = n$_{\rm NLTE}$/n$_{\rm LTE}$, in
the 4600/1.40/$-2.57$ model atmosphere calculated with and without collisions with H~I atoms. Here, n$_{\rm NLTE}$ an n$_{\rm LTE}$ are the non-LTE and the Boltzmann-Saha level populations. As would be
expected, including the collisions with H~I atoms reduces the departures from LTE. Despite the fact
that Ba~II is the dominant ionization stage, in the case of purely electronic collisions, the b-factors of even the ground and lower excited levels differ noticeably from unity in the Ba~II line formation region, log~$\tau_{5000}$ = 0 to $-1$, and the b-factors of the \eu{5d}{2}{D}{}{3/2} and \eu{5d}{2}{D}{}{5/2} fine-splitting sublevels begin to diverge already in deep layers (log~$\tau_{5000} \sim 0$).

\begin{table}
	\caption{Atomic parameters of the investigated Ba~II lines.} 
	\label{tab:listba} 
\begin{center}
	\begin{tabular}{ccrcrc}\hline\hline \noalign{\smallskip} 
		$\lambda$ & \Eexc & log $gf$ & log$\Gamma_6$ & \multicolumn{2}{c}{Ref.}  \\
		\cline{5-6}
		(\AA)    & (eV)  &    &  & $gf$  & $\Gamma_6$      \\
		\noalign{\smallskip} \hline \noalign{\smallskip} 
		Ba~II 4554.03 & 0.00 & 0.17 & $-$7.732 & \cite{Gallagher_ba} & \cite{Mashonkina2006} \\
		Ba~II 5853.67 & 0.60 & $-$1.01 & $-$7.584 & \cite{Gallagher_ba} & \cite{1998MNRAS.300..863B}  \\
		Ba~II 6496.90 & 0.60 & $-$0.38 & $-$7.584 & \cite{Gallagher_ba} & \cite{1998MNRAS.300..863B}  \\
		Eu~II 4129.72 & 0.00 &    0.22 &  $-$7.870 & \cite{Lawler_eu} &  \\	
		\noalign{\smallskip}\hline 
	\end{tabular}
\end{center}
\end{table}

\subsection{Synthetic-Spectrum Computations}

The barium and europium abundances are determined by the synthetic-spectrum method, i.e., by fitting the theoretical line profile to the observed one. We use the SynthV code (Tsymbal et al. 2019) together with BinMag\footnote{http://www.astro.uu.se/\~{}oleg/binmag.html}. The b-factors needed to compute the theoretical non-LTE spectra are calculated with the DETAIL code. The list of lines, atomic data, and their sources are given in Table~2. The Eu~II 4129\,\AA\ line is insensitive to the pressure effects, and we adopted log~$\Gamma_6 = -7.870$, by analogy with the Fe~II lines.

For a solar mixture of isotopes the Ba~II 4554\,\AA\ resonance line has 15 components. Since the isotopes $^{134}$Ba and $^{136}$Ba are not produced in the $r$-process, the line consists of 13 components. We use the wavelengths and relative intensities of the components as in our previous studies (see Table~1 in Mashonkina and Zhao, 2006). In our calculations for a different mixture of even and odd isotopes, we treat the change in their relative abundance as a change in the oscillator strengths of the components. For Ba~II 5853 and 6497\,\AA\ the HFS effect is very small: our calculations with and without HFS give a difference in abundance of less than 0.01~dex. Therefore, we use the set of components and their atomic parameters from the VALD database
(Ryabchikova et al. 2015) everywhere for these lines.

The model atmospheres were obtained by interpolation for given \Teff/log~$g$/[Fe/H] in the
MARCS\footnote{\tt http://marcs.astro.uu.se} grid of models (Gustafsson et al. 2008).
The interpolation algorithm from the MARCS site was used.

\subsection{The Influence of Non-LTE on the Barium Abundance Determination}

The effect from the application of the accurate rate coefficients for collisions with H~I was first checked for the solar Ba~II 5853 and 6497\,\AA\ lines. The spectrum of the Sun as a star was taken from the atlas of Kurucz et al. (1984). The MARCS model atmosphere with \Teff\ = 5777~K and log~$g$ = 4.44 is used. We adopt $\xi_t$ = 0.9~\kms. Under the LTE assumption we get $\eps{}$ = 2.34 and 2.39 for Ba~II 5853 and 6497\,\AA, respectively. This is greater than the meteoritic barium abundance, $\eps{met}$ = 2.21$\pm$0.04 (Lodders et al. 2009). The standard abundance scale, in which $\eps{H}$ = 12, is used here. Non-LTE leads to a strengthening of the Ba~II lines and a reduction in the deduced abundance. When only the electronic collisions are taken into account, $\eps{}$ = 2.21 (Ba~II 5853\,\AA) and 2.23 (Ba II 6497\,\AA), which is consistent with $\eps{met}$. Note that we obtained the same results previously (Mashonkina et al. 1999). Allowance for the collisions with H~I causes the departures from LTE to decrease, and we get a slightly higher abundance, $\eps{}$ = 2.25 and 2.26, although the discrepancies with the meteoritic one do not exceed the determination error. The theoretical non-LTE profiles that describe best the solar Ba~II lines are shown in Fig.~3.

\begin{figure*}  %[htbp]
\includegraphics[width=80mm]{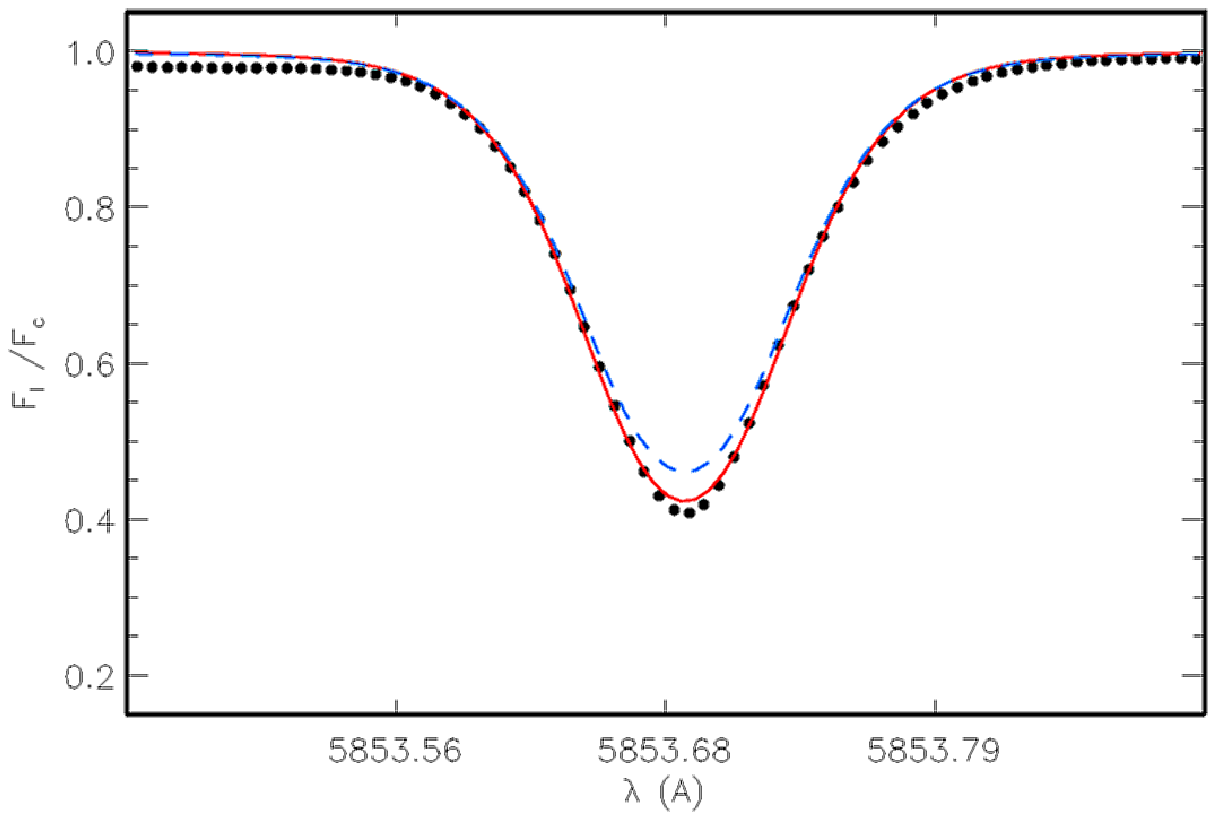}
\includegraphics[width=80mm]{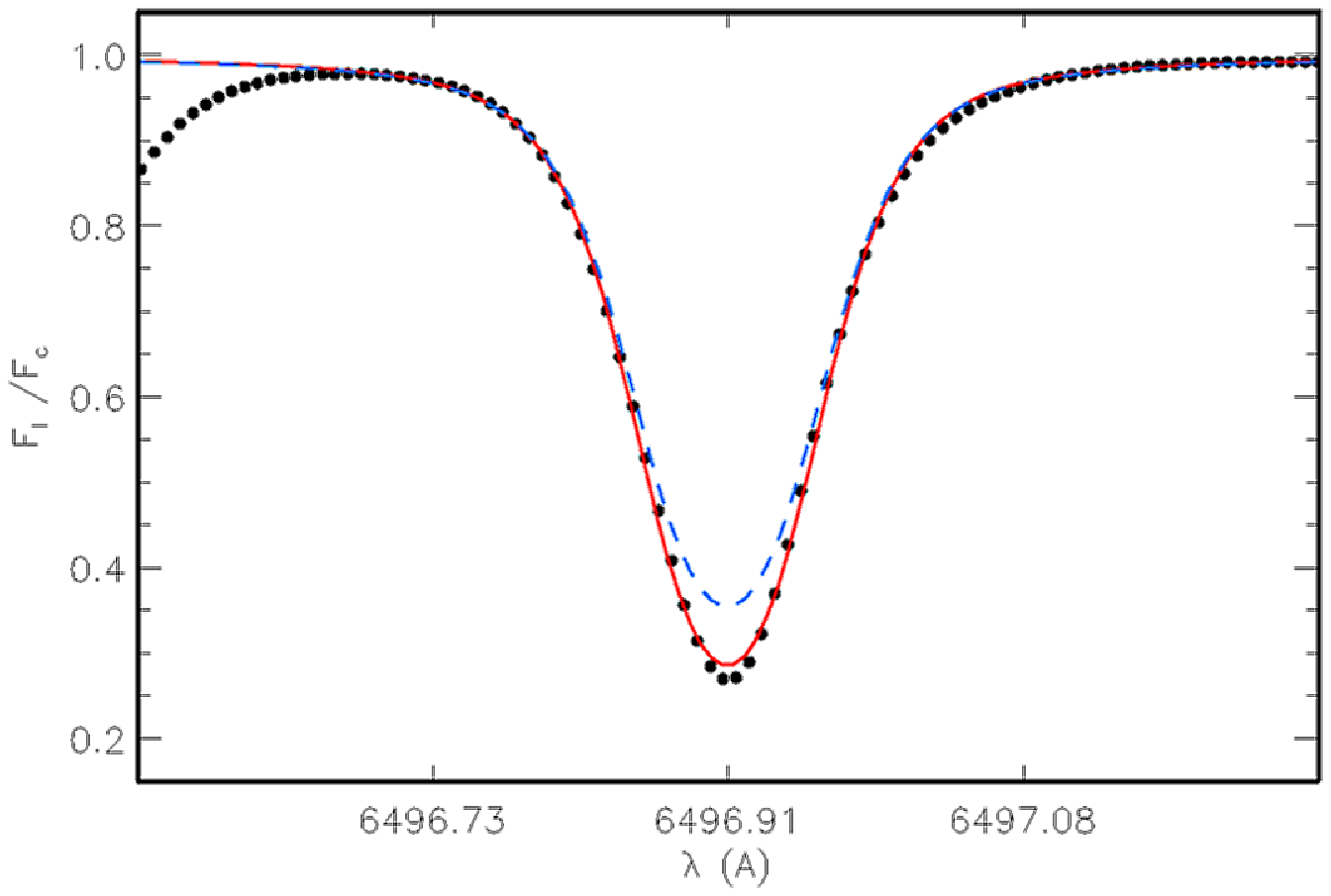}
\caption{Ba II lines in the solar spectrum (dotted curves) and the theoretical non-LTE (solid curves) and LTE
(dashed curves) profiles computed with $\eps{}$ = 2.25 and 2.26 for Ba~II 5853 and 6497~\AA, respectively.}\label{fig:solar}
\end{figure*}

We calculated the non-LTE abundance corrections, $\Delta_{\rm NLTE} = \eps{NLTE} - \eps{LTE}$, for five Ba~II
lines, three from Table~2 as well as Ba~II 4934 and 6141\,\AA, and for two grids of model atmospheres. One
of them has a range of parameters typical for cool giants: \Teff\ = 4500 and 4750~K, log~$g$ from 0.5 to 2.5
with a 0.5 step, and [Fe/H] = 0, $-1$, $-2$, $-2.5$, $-3$. The other one has a range of parameters typical for stars of late spectral types near the main sequence: \Teff\ from 4500 to 6500~K with a step of 500~K, log~$g$ from 3.0 to 4.5 with a 0.5 step, and [Fe/H] from 0 to $-3$ with a $-0.5$ step. The tables of corrections
are accessible at {\tt http://www.inasan.rssi.ru/$\sim$lima/NLTE\_corrections/}.

The non-LTE corrections for three lines at two values of \Teff\ are presented in Fig.~4. As was first
shown by Mashonkina et al. (1999), the departures from LTE for Ba~II lines can have different signs and
values, depending on the stellar metallicity, \Teff, and log~$g$. The picture does not change qualitatively in the calculations with more accurate collisional data either, as can be seen from Fig.~4. At a solar metallicity non-LTE leads to a strengthening of the Ba~II lines and a negative $\Delta_{\rm NLTE}$, irrespective of \Teff\ and log~$g$. We note that the non-LTE effects are small for Ba~II 4554\,\AA\ in cool giants in a wide
metallicity range, because this line is very strong, with well-developed van der Waals wings forming in
deep atmospheric layers. In models with log~$g \ge 3$, the non-LTE effects grow, i.e., $\Delta_{\rm NLTE}$ becomes more negative, as [Fe/H] decreases to a certain value that depends on the line, log~$g$, and \Teff. As [Fe/H] decreases further, the non-LTE corrections are reduced in absolute value, pass through 0, and become
positive. This is because the lines weaken and their formation depths are shifted into deep atmospheric
layers. When the Ba~II line is strong, the non-LTE effect takes place in its core and it is attributable to
dropping the line source function relative to the Planck function in the surface layers due to the escape of photons in the line itself, which leads to a depopulation of the upper level relative to the lower one. When the Ba~II line is weak, it is formed in deep layers, where the upper levels are overpopulated relative to the lower ones, which leads to a line weakening compared to the LTE case.

\begin{figure*}  %[htbp]
	\hspace{-6mm}
\includegraphics[width=50mm]{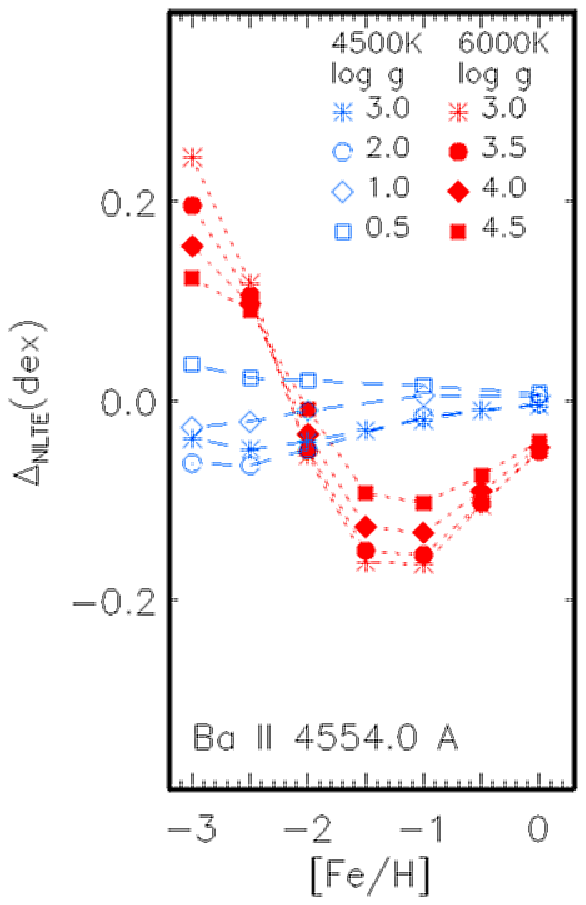}
%\hspace{-7mm}
\includegraphics[width=50mm]{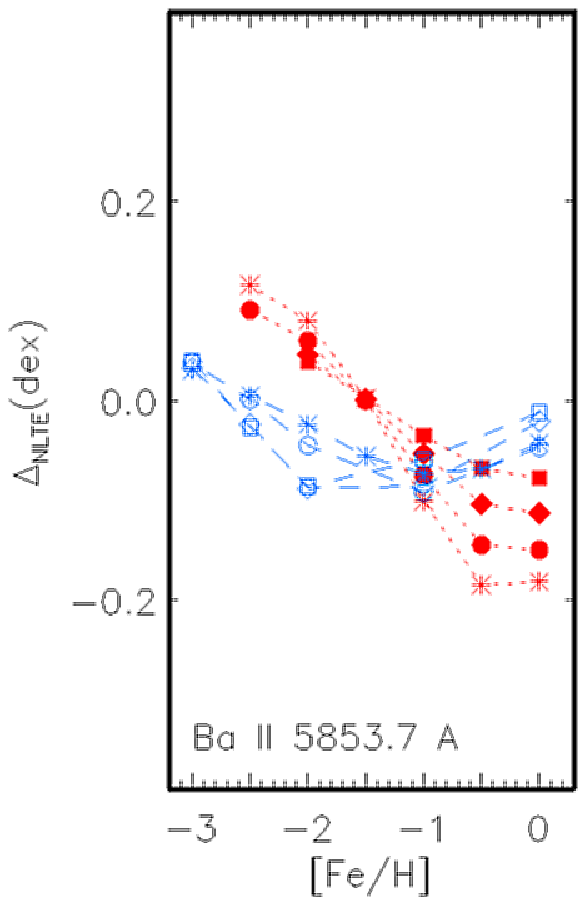}
%\hspace{-7mm}
\includegraphics[width=50mm]{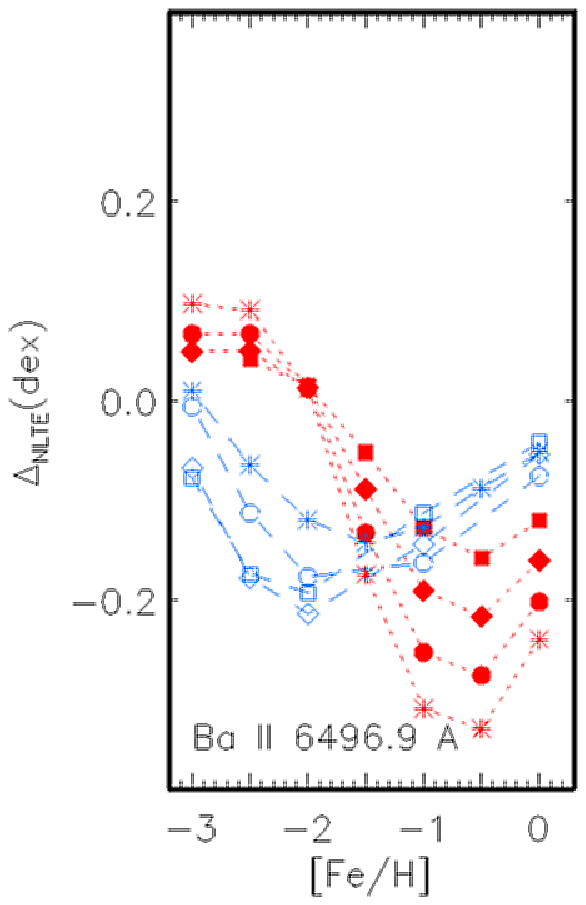}
\caption{Non-LTE abundance corrections for the Ba~II 4554,  5853 and 6497~\AA\ lines as a function of
metallicity and surface gravity. \lgg takes values of 3.0, 3.5, 4.0, 4.5 in the models with \Teff\ = 6000~K and 0.5, 1.0, 2.0, 3.0 in the models with \Teff\ = 4500~K. The non-LTE corrections are not shown if the line equivalent width falls below 3~m\AA. }\label{fig:dnlte}
\end{figure*}

\section{THE BARIUM AND EUROPIUM ABUNDANCES OF THE PROGRAM STARS}\label{Sect:Abund}

The Ba~II lines in the spectra of the program stars are shown in Fig.~5. The procedure for matching the
observed and theoretical profiles leads to an uncertainty in the abundance of no more than 0.02~dex. For
Ba~II 4554\,\AA\ the calculations were carried out with  $f_{\rm odd}$ = 0.18, 0.30, 0.35, and 0.46. The results of our LTE and non-LTE abundance determinations are presented in Table~3, except the abundance for $f_{\rm odd}$ = 0.35.

\begin{table*}
	\caption{LTE and non-LTE abundances from the Ba~II lines at a given fraction of odd isotopes ($f_{\rm odd}$). The uncertainties in the abundance due to the uncertainty in the atmospheric parameters are given for each line.} 
	\label{tab:lines} 
\begin{center}
	\begin{tabular}{lccccc}\hline\hline \noalign{\smallskip} 
$\lambda$, \AA & 5853 & 6496 & \multicolumn{3}{c}{4554} \\
 \cline{4-6} 
$f_{\rm odd}$ & 0.46 &  0.46 & 0.18 & 0.30 & 0.46  \\
\hline\hline
\multicolumn{6}{c}{HD 2796}  \\
$\eps{}$, LTE &  -0.42 &  -0.25 &  -0.08 &  -0.21 &  -0.32  \\
\cline{2-3} 
Mean & \multicolumn{2}{c}{$-0.34\pm0.12$} & & & \\
$\Delta_{\rm NLTE}$ &   0.02 &  -0.11 &  -0.09 &  -0.09 &  -0.10 \\
 $\eps{}$, non-LTE &  -0.40 &  -0.36 &  -0.17 &  -0.30 &  -0.42  \\
\cline{2-3} 
Mean  & \multicolumn{2}{c}{$-0.38\pm0.03$} & & & \\
\multicolumn{6}{c}{Uncertainties in abundance (dex)} \\
\Teff, 100~K &   0.06 &   0.07 &   0.10 &   0.10 &   0.10  \\
\lgg, $-0.05$ & -0.02 &  -0.02 &  -0.01 &  -0.01 &  -0.01  \\
$\xi_t$, 0.1~\kms & -0.01 &  -0.05 &  -0.09 &  -0.09 &  -0.09 \\
\hline\hline
 \multicolumn{6}{c}{HD 108317} \\
$\eps{}$, LTE &   -0.05 &   0.01 &   0.39 &   0.27 &   0.17 \\
\cline{2-3} 
Mean &  \multicolumn{2}{c}{$-0.02\pm0.04$} & & & \\
$\Delta_{\rm NLTE}$ &   0.02 &  -0.08 &  -0.14 &  -0.15 &  -0.17 \\ 
$\eps{}$, non-LTE &  -0.03 &  -0.07 &   0.25 &   0.12 &   0.00 \\
\cline{2-3} 
Mean  & \multicolumn{2}{c}{$-0.05\pm0.03$} & & & \\
\multicolumn{6}{c}{Uncertainties in abundance (dex)} \\
\Teff, 100~K &  0.06 &   0.07 &   0.11 &   0.11 &   0.11 \\
\lgg, $-0.03$ &  -0.01 &  -0.01 &   0.00 &   0.00 &   0.00 \\
$\xi_t$, 0.1~\kms &  -0.01 &  -0.04 &  -0.07 &  -0.07 &  -0.07 \\
\hline\hline
\multicolumn{6}{c}{HD 122563} \\              
$\eps{}$, LTE &  -1.45 &  -1.36 &  -1.22 &  -1.33 &  -1.44   \\
\cline{2-3} 
Mean & \multicolumn{2}{c}{$-1.40\pm0.06$} & & &  \\
$\Delta_{\rm NLTE}$ & 0.07 &   0.03 &   0.04 &   0.04 &   0.04 \\
 $\eps{}$, non-LTE &  -1.38 &  -1.33 &  -1.18 &  -1.29 &  -1.40  \\
\cline{2-3} 
Mean & \multicolumn{2}{c}{$-1.36\pm0.04$} & & & \\
\multicolumn{6}{c}{Uncertainties in abundance (dex)} \\
\Teff, 50~K &   0.05 &   0.05 &   0.05 &   0.05 &   0.05  \\
\lgg, $-0.03$ &  -0.02 &  -0.01 &  -0.01 &  -0.01 &  -0.01 \\
$\xi_t$, 0.1~\kms &  -0.00 &  -0.02 &  -0.06 &  -0.06 &  -0.06  \\
\hline\hline
\multicolumn{6}{c}{HD 128279} \\ 
$\eps{}$, LTE &  -0.49 &  -0.43 &  -0.25 &  -0.39 &  -0.51 \\
\cline{2-3} 
Mean & \multicolumn{2}{c}{$-0.46\pm0.04$} & & & \\
$\Delta_{\rm NLTE}$ &  0.04 &  -0.01 &  -0.08 &  -0.07 &  -0.06 \\
$\eps{}$, non-LTE &  -0.45 &  -0.44 &  -0.33 &  -0.46 &  -0.57 \\
\cline{2-3} 
Mean &  \multicolumn{2}{c}{$-0.45\pm0.01$} & & & \\
\multicolumn{6}{c}{Uncertainties in abundance (dex)} \\
\Teff, 100~K &   0.06 &   0.06 &   0.10 &   0.10 &   0.10 \\
\lgg, $-0.03$ &  -0.00 &   0.01 &  -0.01 &  -0.01 &  -0.01 \\
$\xi_t$, 0.1~\kms &  -0.00 &  -0.02 &  -0.05 &  -0.05 &  -0.05 \\
\hline
\end{tabular}
\end{center}
\end{table*}

\begin{figure*}  %[htbp]
\includegraphics{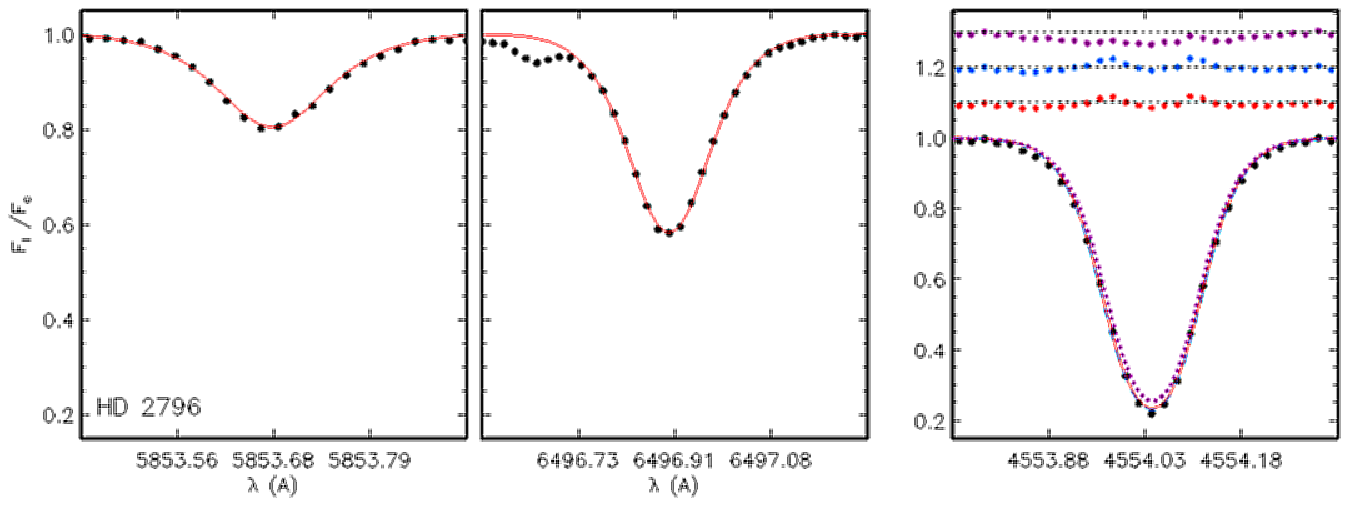}
\includegraphics{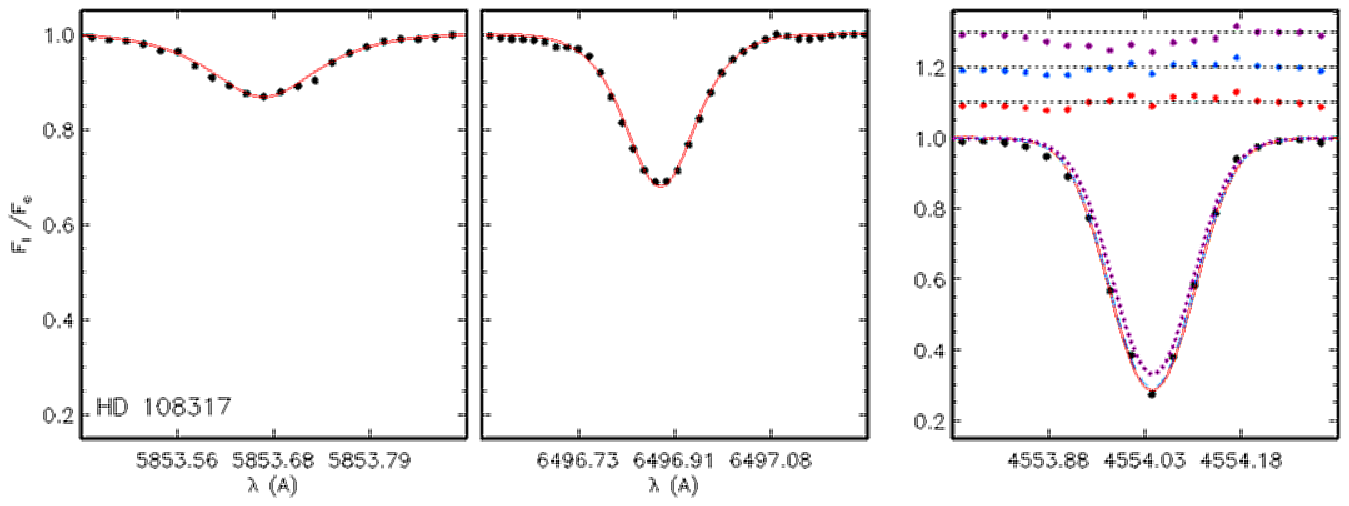}
\includegraphics{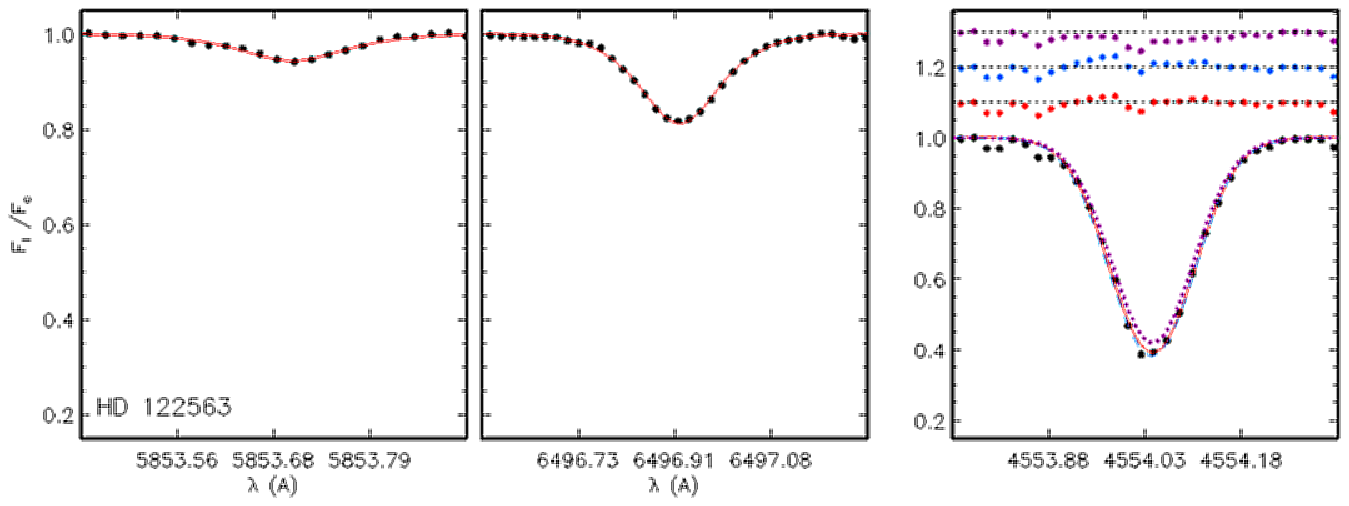}
\includegraphics{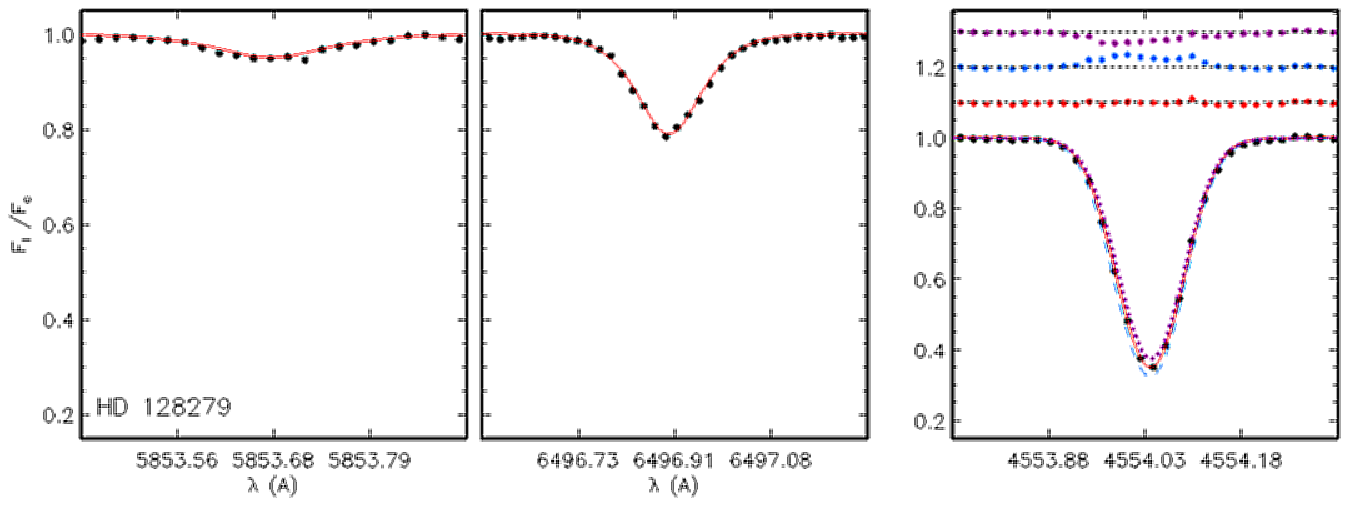}
	\caption{Ba II lines in the stellar spectra (dotted curves) and the theoretical non-LTE spectra (solid curves) computed with the abundances from Table~3. The calculations of Ba~II 4554~\AA\ were made with $f_{\rm odd}$ = 0.18 (lilac dotted curves), 0.46 (blue dashed curves), and the value derived for a given star (red solid curves), as indicated in Table~4. The corresponding O-C values are shown in the upper part of each panel for Ba~II 4554~\AA. }\label{fig:lines}
\end{figure*}

For the star HD 122563 with the lowest metal abundance in our sample non-LTE leads to a weakening of all three Ba~II lines and positive abundance corrections (Table~3). In the remaining three stars, $\Delta_{\rm NLTE} > 0$ only for the weakest line, Ba~II 5853~\AA, while for the other two lines non-LTE leads to their strengthening and $\Delta_{\rm NLTE} < 0$.

For each star the difference in the abundance deduced from the two subordinate lines is smaller in the non-LTE calculations than in the LTE ones. We estimate the accuracy of determined total barium abundance by the root-mean-square error, $\sigma = \sqrt{\Sigma(\overline{x}-x_i)^2/(n-1)}$, where $n$ = 2 is the number of subordinate lines. In the non-LTE calculations, $\sigma$ does not exceed 0.04~dex.

The HFS components formed by the isotopes $^{135}$Ba and $^{137}$Ba increase the full width at half maximum (FWHM) of the Ba~II 4554~\AA\ line. An increase in $f_{\rm odd}$ leads to a growth in the FWHM and absorption and a reduction in the abundance from this line. For the program stars the difference in abundance between $f_{\rm odd}$ = 0.18 and 0.46 is between 0.22 and 0.25~dex. This is considerably larger than the error in the total barium abundance, which makes it possible to determine $f_{\rm odd}$ from the requirement that the abundances from the resonance and subordinate lines be equal.

To compare two indicators of $r/s$, $f_{\rm odd}$ and [Eu/Ba], the abundance was also determined for europium. The non-LTE calculations were performed with the model atom from Mashonkina et al. (2000). Due to the absence of accurate data for collisions with H I, we use the Drawinian rates scaled by a factor of \kH\ = 0.1. Non-LTE leads to a weakening of the Eu~II 4129~\AA\ line and a higher europium abundance, by 0.07-0.1~dex
for different stars. Europium is represented in nature mainly by two isotopes, $^{151}$Eu and $^{153}$Eu, and each of them forms 16 HFS components in the Eu~II 4129~\AA\ line. The $s$-process calculations and the “solar” $r$-process based on them predict that the relative yield of
europium isotopes is similar in the  $s$- and $r$-processes (see, e.g., Travaglio et al. 1999; Bisterzo et al. 2014) and close to the solar abundance ratio: $^{151}$Eu : $^{153}$Eu = 47.8 : 52.2 (Lodders et al. 2009). The set of Eu~II 4129~\AA\ line components and their $gf$-values corresponding to a solar mixture of isotopes were taken from the VALD database. The results of our abundance determinations are presented in Table~4.

\begin{table*}
	\caption{Non-LTE barium and europium abundances and the fraction (in \%) of odd barium isotopes}\label{tab:abund}
\begin{center}
	\begin{tabular}{rccccc}
		\hline\hline
\multicolumn{1}{l}{HD}  & [Fe/H] & $\eps{Ba}$     & $\eps{Eu}$ & [Eu/Ba] & $f_{\rm odd} \times$ 100 \\
\hline      
		2796   &  $-2.19$ & $-0.38\pm0.03$ & $-1.73$ & 0.31 & 40$^{51}_{29}\pm 5$  \\
		108317 &  $-2.24$ & $-0.05\pm0.03$ & $-1.11$ & 0.60 & 53$^{61}_{45}\pm 4$ \\
		122563 &  $-2.55$ & $-1.35\pm0.04$ & $-2.70$ & 0.31 & 39$^{45}_{33}\pm 6$ \\
		128279 &  $-2.19$ & $-0.43\pm0.03$ & $-1.65$ & 0.44 & 27$^{35}_{21}\pm 1$  \\
\hline
\end{tabular}
\end{center}
\end{table*}

{\bf Uncertainties in the derived barium abundances.} We analyzed the influence of uncertainties in the atmospheric parameters on the abundance deduced from the resonance and subordinate lines. The errors in the effective temperature were taken from the original sources. The errors in log~$g$ were calculated by taking into account the distance errors, as given by Bailer-Jones et al. (2018). We estimate the microturbulence error as 0.1~\kms. The results are presented in Table~3. The influence of the error in log~$g$ on the abundance is virtually the same for different lines and it is small. The Ba~II 4554\,\AA\ line with $W_{obs}$ in the range from 80 to 140~m\AA\ lies on the saturation part of the curve of growth and, therefore, it is more sensitive to the errors in \Teff\ and $\xi_t$ than the subordinate lines. Note that for an individual star the abundance errors due to the uncertainty in a given atmospheric parameter are not statistical, but systematic in nature for different lines. For example, for HD~2796 an error of 100~K in \Teff\ leads to a difference of 0.04~dex in abundance between the resonance and subordinate lines, while an error of 0.1~\kms\ in $\xi_t$ leads to a difference of 0.06~dex in abundance. Since the errors in \Teff\ and $\xi_t$ are uncorrelated, the total abundance error due to the errors in the atmospheric parameters will be $\sigma_{atm}$ = 0.07~dex.

The uncertainty in the data for Ba~II + H~I collisions can also be a source of errors in the abundance. An upper limit for this error can be estimated by comparing the abundances deduced in the non-LTE calculations with and without collisions with H~I. This error is also systematic in nature, because ignoring the collisions with H~I leads to an enhancement of the departures from LTE for both resonance and subordinate lines, but to a different extent. For example, for HD~122563 allowance for only the electronic collisions leads to an increase in
the abundance compared to what is obtained when the collisions with H~I are taken into account, by 0.07~dex from Ba~II 4554\,\AA\ and by 0.06~dex and 0.02~dex from Ba II 5853 and 6497\,\AA. In other stars, the shift in the abundance from the Ba~II 4554, 5853, and 6497\,\AA\ lines is: +0.05,+0.05, and +0.05 dex (HD~2796); +0.05,+0.02, and $-0.01$~dex (HD~108317); and +0.02,+0.03, and +0.02 dex (HD~128279).

\section{DETERMINATION OF $f_{\rm odd}$ AND DISCUSSION OF RESULTS}
\label{sect:fodd}

For each star $f_{\rm odd}$ was derived from the requirement that the abundances determined from the resonance
and subordinate lines be equal. The values of $f_{\rm odd}$ are given in Table~4. The statistical error in $f_{\rm odd}$ is determined by the error in the total barium abundance ($\sigma$). We also provide the upper and lower
limits for $f_{\rm odd}$ estimated by taking into account the errors in the atmospheric parameters ($\sigma_{atm}$). Let us give an example for the star HD~2796. We obtained $\eps{NLTE} = -0.38$ from the subordinate lines, while Ba~II 4554\,\AA\ gives $\eps{NLTE} = -0.35$ and $-0.42$ at $f_{\rm odd}$ = 0.35 and 0.46. By assuming that
in this interval $\eps{}$ depends linearly on $f_{\rm odd}$, we get $f_{\rm odd}$ = 0.40. Since $\sigma$ = 0.03~dex, the statistical error in $f_{\rm odd}$ is $\pm 0.05$. The errors in the atmospheric parameters give $\sigma_{atm}$ = 0.07~dex. A shift in the total barium abundance by $+-0.07$~dex implies a change in $f_{\rm odd}$ by $-+0.11$ for HD~2796.

Changing the line formation scenario or, more specifically, passing to allowance for the collisions only with electrons does not affect the determination of $f_{\rm odd}$ for HD~2796 and HD~128279, because this shifts the abundances from the resonance and subordinate lines by the same amount. For the other two stars the abundance from the resonance line increases by a larger value than the one from the subordinate lines. For each of them this implies that $f_{\rm odd}$ can be larger than its value in Table~4, by 0.05.

For three stars we obtained $f_{\rm odd} \ge$ 0.4. This points to a significant or even dominant contribution of
the $r$-process to the observed barium abundances. For the $r$-process Travaglio et al. (1999), Kratz et al. (2007), and Bisterzo et al. (2014) predicted $f_{\rm odd,r}$ = 0.46, 0.44, and 0.60, respectively. If we adopt $f_{\rm odd,r}$ = 0.46, then $f_{\rm odd}$ = 0.4 observed in the star implies that 83~\%\ of its barium was produced in
the $r$-process. The dominance of the $r$-process at the formation epoch of our objects is independently confirmed by the observed europium overabundance relative to barium, with [Eu/Ba] $> 0.3$. The same authors predicted [Eu/Ba]$_r$ = 0.67, 0.63, and 0.80 for the $r$-process. We note the large scatter of predicted values and the agreement of the observations only with the lower boundary of the predictions. In HD~128279, $f_{\rm odd}$ = 0.27 exceeds the solar value, though insignificantly, while [Eu/Ba] = 0.44 suggests a substantial contribution of the $r$-process to the barium abundance.

Of the four stars, $f_{\rm odd}$ was determined previously only for HD~122563, but a lower value of 0.22$\pm$0.15 was obtained by Mashonkina et al. (2008). The previous paper differs from this one in that other
values of log~$g$ = 1.5 and $\xi_t$ = 1.9~\kms\ were used and that the collisions with H~I were disregarded. Our analysis showed that the difference in $\xi_t$ is mainly responsible for the difference in $f_{\rm odd}$. Mashonkina et al. (2008) obtained a higher microturbulence velocity, because they used only Fe~II lines, which have a limited range of equivalent widths in a star with [Fe/H] $\simeq -2.5$, while an efficient non-LTE method
has not yet been developed for Fe~I lines. Thus, we confirm that the method of determining $f_{\rm odd}$ used in
our paper is very sensitive to the accuracy of determining the atmospheric parameters, including $\xi_t$.

Among several works on the determination of $f_{\rm odd}$ by a method that differs from our one and is based on
the dependence of the FWHM of the Ba~II 4554\,\AA\ line on the abundance of odd isotopes, we will allude
only to Gallagher et al. (2015), where the line profile was computed in a 3D model atmosphere. For the
halo star HD~140283 ([Fe/H] $\simeq -2.4$) they obtained $f_{\rm odd}$ = 0.38$\pm$0.02 with a systematic error of 0.06.

We checked how using accurate rate coefficients for the collisions with H~I affects the determination of $f_{\rm odd}$ for two stars in the Sculptor dwarf spheroidal galaxy, ET0381 and 03\_059. Jablonka et al. (2015) used the same method as that in this paper, except for the non-LTE calculations in which the collisions with
H~I were taken into account with the Drawinian rates reduced by a factor of 100, and obtained $f_{\rm odd} \le$ 0.11. Such low values point to the dominance of the $s$-process at the epoch when the stars were formed. However, this is a very strange result, because both stars are metal-poor, with [Fe/H] = $-2.19$ and $-2.88$, and they are expected to have been formed before the first AGB stars appeared in the dwarf galaxy and nucleosynthesis in the $s$-process began. Jablonka et al. (2015) convincingly proved that the existing uncertainties in the atmospheric parameters could not be responsible for the low $f_{\rm odd}$. Later on, the atmospheric parameters of both stars were carefully double checked (Mashonkina et al. 2017). The non-LTE calculations with accurate rate coefficients for Ba~II + H~I collisions confirmed the previously published $f_{\rm odd}$ for the stars ET0381 and 03\_059. For the first star the abundance from the subordinate lines decreased by 0.015~dex and the one from the Ba~II 4554\,\AA\ line did not change. For the second star new calculations led to the same decrease in the abundance from all Ba~II lines, by 0.03~dex. The [Eu/Ba] ratios cannot be determined in these stars, because it is impossible to measure the Eu~II lines. Thus, the question about the source of barium in the Sculptor dwarf spheroidal galaxy at the formation epoch of very metal-poor stars remains unsolved.

\section{CONCLUSIONS}\label{sect:conclusions}

The Ba~II model atom developed previously (Mashonkina et al. 1999) was updated by taking into account the H~I impact excitation and using the rate coefficients from the quantum-mechanical calculations of Belyaev and
Yakovleva (2018). As expected, the non-LTE calculations with this model atom lead to a reduction in the
departures from LTE compared to the case where the collisions only with electrons are taken into account.
With the new model atom we calculated the non-LTE abundance corrections for five Ba~II lines and investigated
their dependence on atmospheric parameters in the ranges \Teff\ = 4500-6500~K, log~$g$ = 0.5-4.5, and [Fe/H] from 0 to $-3$. The tables of corrections are accessible at {\tt http://www.inasan.rssi.ru/$\sim$lima/NLTE\_corrections/}.

To determine the fraction of barium isotopes with an odd mass number, we chose four Galactic halo giants for which there are high-resolution and high S/N spectra and reliable atmospheric parameters, namely the effective temperature and surface gravity, obtained from non-spectroscopic methods based, among other things, on the Gaia
parallaxes (Gaia Collaboration 2018). We used a method based on the requirement that the abundances from the Ba~II 4554\,\AA\ resonance line and the Ba~II 5853 and 6497\,\AA\ subordinate lines be equal. A combination of accurate atmospheric parameters, high-quality observed spectra, and theoretical spectra computed with the most complete accounting for the physical processes in the Ba~II atom provides a high accuracy of determining the abundance from individual lines. Good agreement between the non-LTE abundances from the two subordinate lines, so that for each star the standard deviation does not exceed 0.04~dex, serves as a confirmation.

For three stars (HD~2796, HD~108317, and HD~122563) $f_{\rm odd} \ge$ 0.4. This suggests that $\ge$80~\%\ of the barium observed in these stars was synthesized in the $r$-process. Our estimate was obtained using $f_{\rm odd,r}$ = 0.46 (Travaglio et al. 1999). In HD~128279, our value of $f_{\rm odd}$ = 0.27 exceeds the fraction of odd
isotopes in the Solar system, but only slightly. The dominance of the $r$-process at the formation epoch of
our sample stars is confirmed by the presence of a europium overabundance relative to barium, with [Eu/Ba] $> 0.3$.

{\it Acknowledgements.} We are grateful to Rana Ezzeddine who provided the spectra of HD~108317. We used the
VLT2/UVES archive of observed spectra and the ADS\footnote{http://adsabs.harvard.edu/abstract\_service.html}, SIMBAD, MARCS, and VALD databases. The work presented in Section~2 was performed as part of RSF project no. 17-13-01144.

\end{document}